\documentclass[
prd
,showpacs,amssymb,superscriptaddress,aps
]{revtex4-2}
\usepackage{graphicx}
\usepackage{color}
\input{epsf}

\usepackage{amsmath,amssymb}
\usepackage{bm}
\usepackage{times}
\usepackage{ulem}

\newcommand{\dalm}{\kern1pt\vbox{\hrule height 0.9pt\hbox{\vrule width 0.9pt
\hskip 2.5pt\vbox{\vskip 5.5pt}\hskip 3pt\vrule width 0.3pt}\hrule height 0.3pt}
\kern1pt}

\newcommand{\gsim}{\, \raisebox{-0.8ex}{$\stackrel{\textstyle >}{\sim}$ }}

\begin{document}



\title{Shear and interface modes in neutron stars with pasta structures}

\author{Hajime Sotani}
\email{sotani@yukawa.kyoto-u.ac.jp}
\affiliation{Astrophysical Big Bang Laboratory, RIKEN, Saitama 351-0198, Japan}
\affiliation{Interdisciplinary Theoretical \& Mathematical Science Program (iTHEMS), RIKEN, Saitama 351-0198, Japan}



\date{\today}

\begin{abstract}
We carefully examine the shear and interface modes, which are excited due to the presence of crust elasticity, in neutron stars with pasta structures, adopting the relativistic Cowling approximation. We find that the shear modes are independent of the presence of the cylindrical-hole and spherical-hole nuclei at least up to a few kilohertz, while the interface modes strongly depend on the presence of the cylindrical-hole and spherical-hole nuclei. In addition, we find empirical relations for the interface mode frequencies multiplied by the stellar mass and for the shear mode frequencies multiplied by the stellar radius. These relations are expressed as a function of the stellar compactness almost independently of the stiffness in a higher-density region inside the neutron star, once one selects the crust equation of state. Thus, if one would simultaneously observe the shear and interface modes from a neutron star, one might extract the neutron star mass and radius with the help of the constraint on the crust stiffness obtained from terrestrial experiments. 
\end{abstract}

\pacs{04.40.Dg, 97.10.Sj, 04.30.-w, 26.60.Gj}
%
\maketitle


\section{Introduction}
\label{sec:I}

Neutron stars are one of the most suitable natural laboratories for probing physics under extreme states. This object is provided as a massive remnant remaining after the supernova explosion, which happens at the last moment of the star's life \cite{ST83}. It is considered that the density inside the neutron stars easily exceeds the standard nuclear density, while the gravitational and magnetic fields inside/around the neutron stars become much stronger than those observed in our Solar system. Thus, one could inversely see the aspect of such extreme conditions by carefully observing the neutron stars and their phenomena. 
    
One of the most important observational pieces of evidence is the discovery of the $2M_\odot$ neutron stars \cite{D10,A13,C20,F21}. It is theoretically known that the neutron star has a maximum mass, which depends on the equation of state (EOS) for neutron star matter. Owing to the discovery of the $2M_\odot$ neutron stars, now one can exclude the EOSs, with which the expected maximum mass does not reach the observations. Meanwhile, through careful observations of the pulsar light curve, one may extract the neutron star properties, especially the stellar compactness. This is because the light emitted from the neutron star's surface can bend due to the strong gravitational field induced by the neutron star, which is one of the relativistic effects (e.g., Refs. \cite{PFC83,LL95,PG03,PO14,SM18,Sotani20a}). In fact, the observations with the Neutron Star Interior Composition ExploreR (NICER), which is a NASA telescope on the International Space Station, give us the constraints on the neutron star mass and radius, i.e., PSR J0030+0451 \cite{Riley19,Miller19} and PSR J0740+6620 \cite{Riley21,Miller21}. Moreover, the observations of gravitational waves from the binary neutron star merger, GW170817 \cite{gw170817}, enables us to constrain the tidal deformability of the neutron stars, which leads to the constraint on the $1.4M_\odot$ neutron star radius \cite{Annala18}. In addition to these astronomical observations, the neutron star properties in a lower-density region can be gradually constrained through terrestrial nuclear experiments. It may be reasonable that one discusses both constraints simultaneously in the neutron star mass and radius plane \cite{SNN22,SO22,SN23}. 

The oscillation frequencies of the neutron stars must be another important piece of information for extracting the neutron star properties. Since the oscillation frequencies strongly depend on the interior properties of the object, one may be able to know the interior properties as an inverse problem by observing the oscillation frequencies. This technique is known as asteroseismology, which is similar to seismology on Earth and helioseismology on Sun. In practice, by identifying the quasi-periodic oscillations observed in the afterglow following the magnetar giant flares \cite{SW2005,SW2006} with the crustal torsional oscillations, one can constrain the crust properties (e.g., Refs. \cite{SW2009,GNHL2011,SNIO2012,SIO2016,SIO2019,SKS23}). This is a good example of how asteroseismology works well. Furthermore, using the gravitational waves from neutron stars, one could get information on neutron star mass, radius, and EOS (e.g., Refs. \cite{AK1996,AK1998,STM2001,SH2003,SYMT2011,PA2012,DGKK2013,KHA2015,Sotani2020,Sotani21,SD2021}). 

In the observation of the neutron star oscillations, several eigenfrequencies may be simultaneously excited, where each eigenfrequency corresponds to a different physical process. By identifying an observed frequency with a specific eigenfrequency one by one, one can see the physics corresponding to the eigenmode. For example, if one would observe the fundamental oscillations of a neutron star, one may extract the average density of the neutron star \cite{AK1996,AK1998}. In a similar way, if one would detect the stellar oscillations associated with the crust elasticity, such as torsional, shear, or interface modes, one may glimpse crust properties. Up to now, there are a few studies about the non-radial oscillations of neutron stars with an elastic crust, e.g., Refs. \cite{Finn90,SL02,PB05,KHA2015}. However, in these previous studies, the shear and interface modes, which are polar-type oscillations exited due to the crust elasticity, have been discussed only on a specific stellar model, where the systematical study has never been done. So, in this study, we will systematically examine the shear and interface modes, adopting the realistic neutron star model with the crust elasticity, where not only the phase composed of spherical nuclei but also the phase composed of non-spherical nuclei, the so-called pasta phase, are considered as crust equilibrium models. As the density increases, the shape of the nuclei changes from spherical (SP) to cylindrical (C), slab-like (S), cylindrical-hole (CH), and spherical-hole (SH) before the matter becomes uniform (U).

This manuscript is organized as follows. In Sec. \ref{sec:EOS}, we mention the equilibrium models together with the EOS adopted in this study and also the shear modulus in the pasta phase. In Sec. \ref{sec:perturbations}, we briefly mention the perturbation equations governing the non-radial oscillations in neutron stars with elastic crust. Then, we discuss the eigenfrequencies in Sec. \ref{sec:modes} and systematically examine their dependence on the neutron star properties in Sec. \ref{sec:dep}. Finally, we conclude this study in Sec. \ref{sec:Conclusion}. Unless otherwise mentioned, we adopt geometric units in the following, $c=G=1$, where $c$ and $G$ denote the speed of light and the gravitational constant, respectively.

\section{EOS and Equilibrium models}
\label{sec:EOS}

In this study, we simply consider a non-rotating, strain-free, and spherically symmetric neutron star as an equilibrium model. The metric describing such an object is given by
\begin{equation}
  ds^2 = -e^{2\Phi}dt^2 + e^{2\Lambda}dr^2 + r^2\left(d\theta^2 + \sin^2\theta d\phi^2\right), \label{eq:metric}
\end{equation}
where $\Phi$ and $\Lambda$ are the metric functions depending on only $r$. In particular, $\Lambda$ is directly connected to the enclosed gravitational mass, $m$, inside the radial position, $r$, through $e^{-2\Lambda}=1-2m/r$. The stellar models are constructed by integrating the Tolman-Oppenheimer-Volkoff equation together with an appropriate EOS for neutron star matter. Since the Fermi temperature of a neutron star is generally much higher than the real temperature of a neutron star, one can neglect the thermal effect on the neutron star structure. But, one may have to carefully handle the surface envelope, where the density becomes low enough for appearing the thermal effect \cite{GPE83}. Nevertheless, in this study, we simply consider the neutron star models, assuming that the surface density is $10^6$ g/cm$^3$ and the density of the surface of the outer crust is $10^{10}$ g/cm$^3$. In addition, in this study, we will consider the shear oscillations, which can be excited due to the non-zero crustal elasticity. Since the shear oscillations are a kind of polar-type oscillations, we have to consider not only the crustal region but also the core region, even though the shear oscillations are confined only inside the crust region. This is a completely different situation from the study of the torsional oscillations (axial-type oscillations), which can be discussed only inside the neutron star crust, e.g., Refs. \cite{SNIO2012,SIO2016}, separately from the neutron star core.

The bulk energy per nucleon for the zero temperature uniform nuclear matter expected for any EOS models is expressed as a function of the baryon number density, $n_{\rm b}$, and an asymmetry parameter, $\alpha$:
\begin{equation}
  \frac{E}{A} = w_s(n_{\rm b}) + \alpha^2 S(n_{\rm b}) + {\cal O}(\alpha^3), \label{eq:E/A}
\end{equation}
where $n_b$ and $\alpha$ are given by $n_{\rm b}=n_n+n_p$ and $\alpha=(n_n-n_p)/n_{\rm b}$ with the neutron number density, $n_n$, and the proton number density, $n_p$. In this expression, $w_s$ corresponds to the energy per nucleon of symmetric nuclear matter ($\alpha=0$), while $S$ denotes the density-dependent symmetry energy. In addition, $w_s$ and $S$ can be expanded in the vicinity of the saturation density, $n_0$, for the symmetric nuclear matter as a function of $u=(n_{\rm b}-n_0)/(3n_0)$:
\begin{gather}
  w_s(n_{\rm b}) = w_0 + \frac{K_0}{2}u^2 + {\cal O}(u^3), \label{eq:ws} \\
  S(n_{\rm b}) = S_0 + Lu + {\cal O}(u^2). \label{eq:S}
\end{gather}
The coefficients in these expansions are the nuclear saturation parameters, which characterize each EOS. That is, each EOS has its own set of nuclear saturation parameters. In this study, to examine the dependence on the symmetry energy, we adopt the phenomenological EOSs constructed by Oyamatsu and Iida \cite{OI03,OI07} (hereafter referred to as OI-EOSs). The OI-EOSs are specially constructed for the given values of $K_0$ and $L$ so that the other saturation parameters are tuned in such a way as to reproduce the empirical nuclear data for stable nuclei, adopting a simplified version of the extended Thomas-Fermi theory \cite{OI03,OI07}. This is because $n_0$, $w_0$, and $S_0$ are well constrained from the terrestrial experiments, while the other parameters, i.e., $K_0$ and $L$ (and the additional saturation parameters associated with the higher terms), are relatively more difficult to be constrained experimentally. Nevertheless, the constraint on $K_0$ is gradually becoming more severe, i.e., $K_0=240\pm 20$ MeV \cite{Sholomo}, while $L$ is still less constrained \cite{Li19,SNN22}. So, in this study, we especially adopt the EOS models with $K_0=230$ MeV. In Table \ref{tab:EOS}, we list the EOS parameters adopted in this study, where we also show the transition density from a specific pasta phase to the next pasta phase. We note that the crust thickness strongly depends on $L$ and stellar compactness, $M/R$, \cite{OI07,SIO2017}, i.e., the thickness decreases as $L$ and $M/R$ increase, and the pasta structures almost disappear, using the EOS model with $L \gsim 100$ MeV~\cite{OI07}.

\begin{table}
\centering
\caption{
The EOS parameters adopted in this study. SP-C, C-S, S-CH, CH-SH, and SH-U denote the transition densities for the OI-EOSs characterized by $K_0$ and $L$. In addition, for the $1.4M_\odot$ neutron star model constructed with each EOS, the ratio of the thickness of the elastic region composed of spherical and cylindrical nuclei, $\Delta R_{\rm SpCy}$, and that composed of cylindrical-hole and spherical-hole nuclei, $\Delta R_{\rm CHSH}$, to the stellar radius, $R$, is also listed.
}
\begin{tabular}{cccccccccc}
\hline\hline
  $K_0$ (MeV)  & $L$ (MeV) & SP-C (fm$^{-3}$) & C-S (fm$^{-3}$) & S-CH (fm$^{-3}$) & CH-SH (fm$^{-3}$) & SH-U (fm$^{-3}$) & $\Delta R_{\rm SpCy}/R$ & $\Delta R_{\rm CHSH}/R$ \\
\hline
  230 & 42.6 & 0.06238 & 0.07671 & 0.08411 & 0.08604   & 0.08637  & 0.0007731 & 0.05406  \\  
  230 & 73.4 & 0.06421 & 0.07099 & 0.07284 & 0.07344   & 0.07345  & 0.0003773 & 0.06740  \\  
\hline\hline
\end{tabular}
\label{tab:EOS}
\end{table}

\begin{figure}[tbp]
\begin{center}
\includegraphics[scale=0.5]{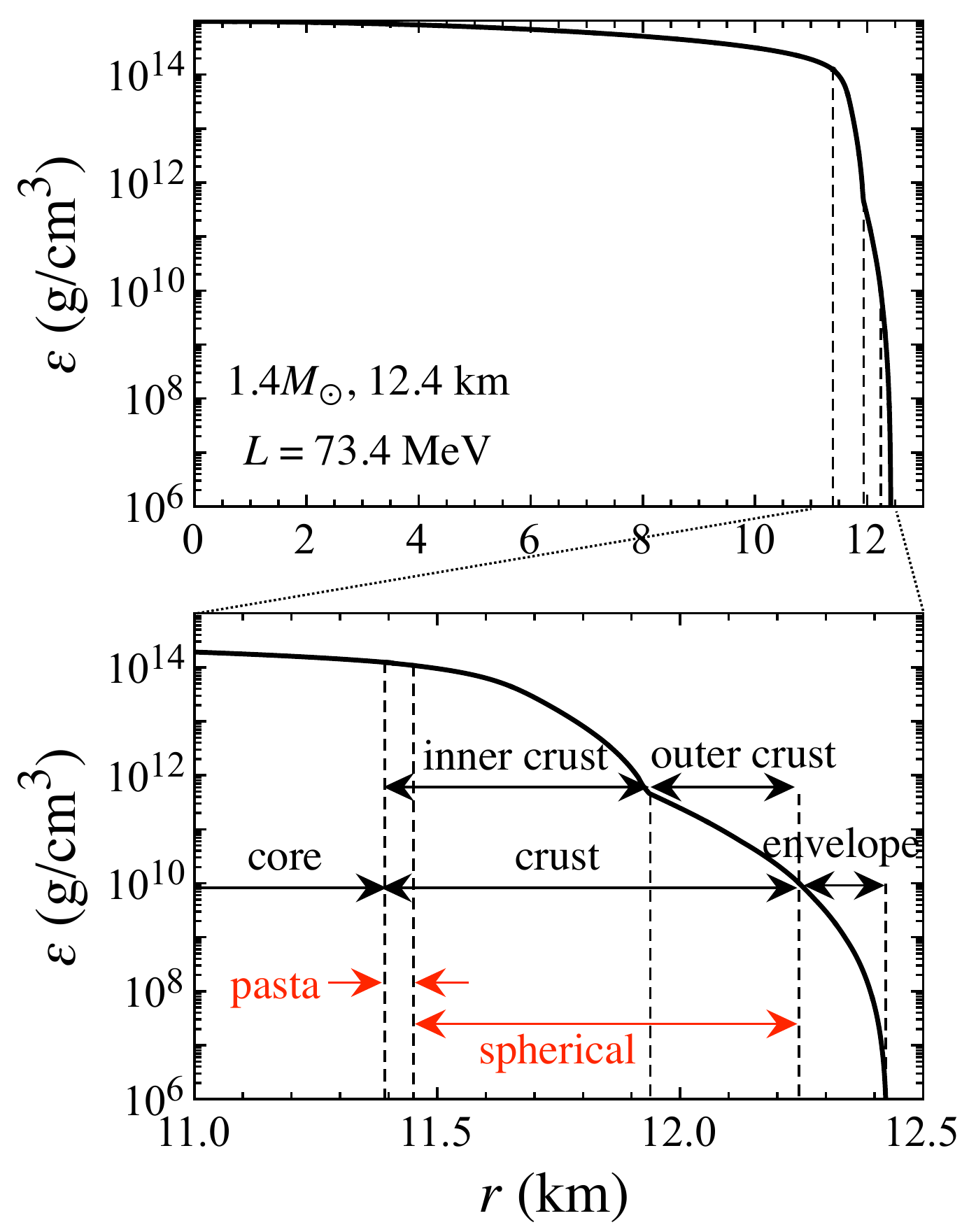}
\end{center}
\caption{
Radial profile of energy density for the neutron star model with $1.4M_\odot$ and 12.4 km, using the EOS with $L=73.4$ MeV. In the top panel, the vertical dashed lines from left to right denote the boundary between the core and crust, the boundary between the inner and outer crust, and the boundary between the crust and envelope. 
}
\label{fig:er}
\end{figure}

\begin{figure}[tbp]
\begin{center}
\includegraphics[scale=0.5]{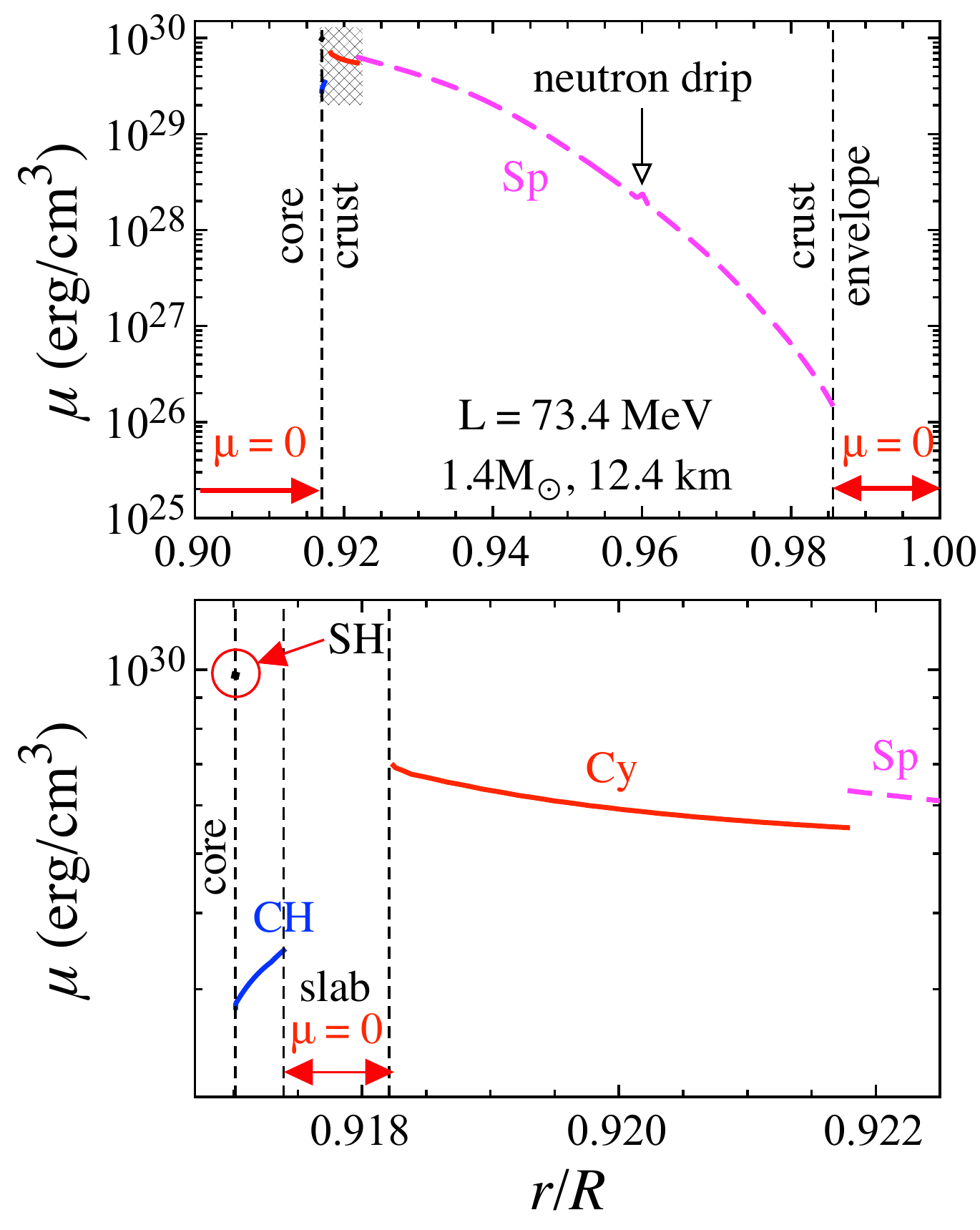}
\end{center}
\caption{
The effective shear modulus for various phases in neutron star curst. The bottom panel is an enlarged view of the shaded region shown in the top panel, where Sp, Cy, CH, and SH denote the phases composed of spherical, cylindrical, cylindrical-hole, and spherical-hole nuclei, respectively. We note that the shear modulus in the phase of slab-like nuclei becomes zero against the linear response \cite{PP1998}. For reference, we show the boundary between the core and crust and the boundary between the crust and envelope in the top panel, and the boundary between the core and SH; the boundary between CH and the phase of slab-like nuclei; and the boundary between the phase of slab-like nuclei and Cy in the bottom panel with the dashed lines.
}
\label{fig:mu}
\end{figure}

The shear modulus, $\mu$, is an additional integrant to discuss the shear oscillations. The shear modulus, $\mu_{\rm sp}$, in the body-centered cubic (bcc) lattice composed of the spherical nuclei has been formulated as a function of the ion number density, $n_i$, the charge number of the ion, $Z$, and a Wigner-Seitz cell radius, $a$, i.e., $4\pi a^3/3 = 1/n_i$~\cite{SHOII1991}:
\begin{equation}
  \mu_{\rm sp} = 0.1194 \frac{n_i(Ze)^2}{a}. \label{eq:mu_sp}
\end{equation}
This expression of the shear modulus should be modified a little due to the phonon contribution~\cite{Baiko2011}, the electron screening effect~\cite{KP2013}, the polycrystalline effect~\cite{KP2015}, and the effect of finite-sizes of atomic nuclei \cite{STT22b}, but in this study, we simply adopt the standard expression given as Eq. (\ref{eq:mu_sp}). On the other hand, the shear modulus, $\mu_{\rm cy}$, in the phase composed of the cylindrical nuclei is expressed as a function of the Coulomb energy per volume of a Wigner-Seitz cell, $E_{\rm Coul}$, and the volume fraction of cylindrical nuclei, $w_2$, as 
\begin{equation}
   \mu_{\rm cy} = \frac{2}{3}E_{\rm Coul}\times 10^{2.1(w_2-0.3)}, \label{eq:mu_cy}
\end{equation}
and the shear modulus, $\mu_{\rm sl}$, in the phase composed of the slab-like nuclei can be considered as 
\begin{equation}
   \mu_{\rm sl} = 0 \label{eq:mu_s}
\end{equation}
against the linear perturbations, i.e., the deformation energy due to the distortion becomes of higher order contribution~\cite{PP1998}. 
We note that the elastic properties of phases with nonspherical nuclei, i.e., pasta phase, in a neutron star have been also discussed in Refs. \cite{CSH18,PZP20}, which suggested the possibility that the tiny but non-zero elastic constant may appear in the polycrystalline lasagna (slab-like nuclei). By taking into account this feature, the results shown in this study may be changed. 
Additionally, the shear modulus, $\mu_{\rm ch}$ ($\mu_{\rm sh}$), in the phase composed of the cylindrical-hole (spherical-hole) nuclei can be derived in the same way as $\mu_{\rm cy}$ ($\mu_{\rm sp}$) because the liquid crystalline structure of cylindrical-hole (spherical-hole) nuclei is the same as that of cylindrical (spherical) nuclei (see Ref. \cite{SIO2019} for details). In Figs.~\ref{fig:er} and \ref{fig:mu}, as an example, we show the radial profile of energy density and shear modulus for the stellar model with $1.4M_\odot$ and $12.4$ km constructed using the OI-EOS with $K_0=230$ and $L=73.4$ MeV.

Finally, we should mention the effect of superfluidity, although we simply neglect such an effect in this study. In general, the enthalpy density effectively decreases because of the fact that the superfluid neutron does not contribute to the oscillations, which leads to an increase in the frequencies (at least for the torsional oscillations)
\cite{SNIO2012,KP18}. We will see how the superfluidity can change the frequencies of the shear and interface modes in the future.

\section{Perturbation equations}
\label{sec:perturbations}

In this study, we simply adopt the relativistic Cowling approximation, i.e., the metric is fixed during the fluid oscillations. Even with this approximation, one can qualitatively discuss the behavior of eigenfrequencies \cite{YK97}. The Lagrangian displacement, $\xi^i$, for the polar-type oscillations is generally given with the spherical harmonics, $Y_{\ell m}(\theta,\phi)$, by 
\begin{equation}
  \xi^i = \left(rW, V \frac{\partial}{\partial\theta}, 
     V\frac{1}{\sin^2\theta}\frac{\partial}{\partial\phi}\right)Y_{\ell m}(\theta,\phi)e^{i\sigma t}, \label{eq:xi} 
\end{equation}
where $W$ and $V$ are the functions of $r$ and $\sigma$ is the eigenvalue, while the pressure perturbation, $\delta p$, is expressed with the energy density, $\epsilon$, and pressure, $p$, for the equilibrium models as 
\begin{align}
  \delta p = (\epsilon + p)H(r)Y_{\ell m}(\theta,\phi)e^{i\sigma t}. \label{eq:dp}
\end{align}
We note that the polar-type oscillations can be discussed completely apart from the axial-type oscillations, because of the nature of the spherically symmetric background. In this study, we also consider the adiabatic oscillations, i.e., 
\begin{equation}
  \Delta p = \frac{p\Gamma}{\epsilon + p}\Delta \epsilon, \label{eq:ad}
\end{equation}
where $\Gamma$ is the adiabatic index and $\Delta Q$ denotes the Lagrangian perturbation of a quantity $Q$, which is associated with the Eulerian perturbation, $\delta Q$, through
\begin{equation}
  \Delta Q = \delta Q + \xi^r \frac{dQ}{dr}. \label{eq:LE}
\end{equation}
So, one can derive the relation between $\delta p$ and $\delta \epsilon$ as
\begin{equation}
  \delta p = c_s^2\delta \epsilon + p\Gamma \xi^r A_r, \label{dp1}
\end{equation}
where $c_s$ is the sound velocity and $A_r$ is the relativistic Schwarzschild discriminant given by
\begin{gather}
  c_s^2 \equiv \left(\frac{\partial p}{\partial \epsilon}\right)_s = \frac{\Delta p}{\Delta\epsilon} 
     =  \frac{p\Gamma}{\epsilon + p}, \label{eq:cs} \\
  A_r = \frac{1}{\epsilon + p}\frac{d\epsilon}{dr} - \frac{1}{p\Gamma}\frac{dp}{dr}.  \label{eq:Ar}
\end{gather}

On the other hand, the shear strain tensor, $\Sigma_{\mu\nu}$, is described through the relation of
\begin{equation}
  {\cal L}_{u}\Sigma_{\mu\nu} = \frac{2}{3}\Sigma_{\mu\nu}\nabla_\alpha u^\alpha 
     + \sigma_{\mu\nu},  \label{eq:Shear}
\end{equation}
where ${\cal L}_u$ is the Lie derivative along the direction of fluid four-velocity, $u^\mu$, and $\sigma_{\mu\nu}$ is the rate of shear tenor \cite{CQ72}, which are respectively expressed as 
\begin{gather}
  {\cal L}_u \Sigma_{\mu\nu} = u^\alpha \nabla_\alpha \Sigma_{\mu\nu} 
     + \Sigma_{\alpha\nu}\nabla_\mu u^{\alpha} + \Sigma_{\mu\alpha} \nabla_{\nu} u^{\alpha}, \\
  \sigma_{\mu\nu} = \frac{1}{2}\left(P^\alpha_{\ \nu}\nabla_\alpha u_\mu
     + P^\alpha_{\ \mu}\nabla_\alpha u_\nu\right) 
     - \frac{1}{3}P_{\mu\nu}\nabla_\alpha u^\alpha. \label{eq:sigma}
\end{gather}
Here, $P_{\mu\nu}$ is the projection tensor given by
\begin{equation}
  P_{\mu\nu} = g_{\mu\nu} + u_\mu u_\nu. \label{eq:projection}
\end{equation}
Using these quantities, the contribution from the shear strain in perturbation of the energy-momentum tensor is given by 
\begin{equation}
  \delta T_{\mu\nu}^{(s)} = -2\mu \delta \Sigma_{\mu\nu}, \label{eq:dts}
\end{equation}
assuming a Hookean relationship \cite{Schumaker83}, where $\mu$ is the shear modulus discussed in the previous section. 

With the relativistic Cowling approximation, the perturbation equations are derived from the linearized energy-momentum conservation laws, i.e., $\nabla_\mu\delta T^{\mu\nu}=0$. The concrete system of equation is shown in Appendix \ref{sec:appendix_1} for the elastic region and in Appendix \ref{sec:appendix_2} for the fluid region ($\mu=0$), while the boundary and junction conditions, which should be imposed, are shown in Appendix  \ref{sec:appendix_3} \cite{SL02}. In practice, the perturbation equations are integrated outward from the center and inward from the stellar surface with the appropriate boundary conditions, where the corresponding solutions are named as $(y_1^{\rm in},y_2^{\rm in})$ and $(y_1^{\rm out},y_2^{\rm out})$, respectively. Then, the eigenfrequencies are determined via the condition of
\begin{equation}
  \Delta \equiv y_1^{\rm in} y_2^{\rm out} - y_1^{\rm out} y_2^{\rm in}=0 \label{eq:Delta}
\end{equation}
at some position inside the star, e.g., the boundary at the crust and envelope.
In this study, we especially focus on the $\ell=2$ modes.

\section{Eigenfrequencies}
\label{sec:modes}

\begin{figure*}[tbp]
\begin{center}
\includegraphics[scale=0.5]{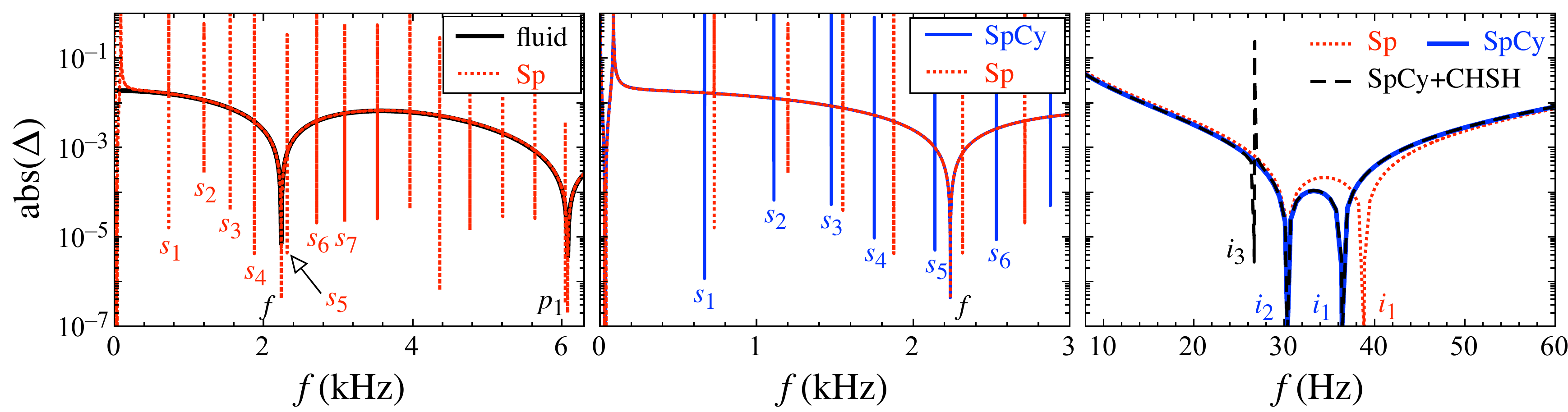}
\end{center}
\caption{
The absolute values of $\Delta$ given by Eq. (\ref{eq:Delta}) are shown as a function of the frequencies. The eigenfrequencies correspond to the specific frequencies, with which the absolute value of $\Delta$ at some position inside the star becomes zero. In the left pane, we show the results for the stellar model composed of only the fluid with the solid line and that including a non-zero elastic region composed of spherical nuclei with the dotted line. In the middle panel, we show the results for the stellar model including a non-zero elastic region composed of spherical nuclei with the dotted line and that including a non-zero elastic region composed of spherical and cylindrical nuclei with the solid line. The right panel is just an enlarged view of the middle panel, where we also show the result for the ``realistic" stellar model with the dashed line. The neutron star model is the same as in Fig.~\ref{fig:mu}.}
\label{fig:delta}
\end{figure*}

In order to understand the dependence of eigenfrequencies excited in the neutron stars on the presence of elasticity, first we consider (i) the neutron star composed of fully zero-elastic ``fluid"; (ii) the stellar model with elastic phase composed of spherical nuclei, ``Sp";  (iii) the stellar model with elastic phase composed of spherical and cylindrical nuclei, ``Sp+Cy"; and (iv) the ``realistic" stellar model with elastic phase composed of spherical, cylindrical, cylindrical-hole, and spherical-hole nuclei as shown in Fig. \ref{fig:mu}. That is, we focus on a specific neutron star model with $1.4M_\odot$ and $12.4$ km constructed with the EOS with $L=73.4$ MeV, but the shear moduli in some elastic phases are artificially put to zero except for the realistic stellar model. In Fig. \ref{fig:delta}, one can see how the eigenfrequencies depend on the elastic phases, where the absolute value of $\Delta$ at the boundary between the crust and envelope is shown as a function of the frequency, i.e., the eigenfrequencies correspond to the frequency where abs$(\Delta)=0$. 
The resultant eigenfrequencies excited in the stellar models shown in Fig. \ref{fig:delta} are listed in Table \ref{tab:modes}.

\begin{table}
\centering
\caption{
Eigenfrequencies excited in the stellar models shown in Fig. \ref{fig:delta} in the unit of kHz.
}
\begin{tabular}{ccccc}
\hline\hline
    & fluid & Sp & SpCy & SpCy+CHSH \\
\hline
  $f$      & 2.237 & 2.237 & 2.237  & 2.237 \\
  $p_1$ & 6.075 & 6.074  & 6.074  & 6.074 \\
  $i_1$  & ---      & 0.039   & 0.036  & 0.036 \\
  $i_2$  & ---      & 0.030   & 0.030  & 0.030 \\
  $i_3$  & ---      & ---      & ---      & 0.027 \\
  $s_1$  & ---     & 0.730   & 0.668  & 0.668  \\
  $s_2$  & ---     & 1.201   & 1.111  & 1.111  \\
  $s_3$  & ---     & 1.552   & 1.477 & 1.477  \\
  $s_4$  & ---     & 1.877   & 1.752  & 1.752  \\  
  $s_5$  & ---     & 2.316   & 2.139  & 2.139  \\
\hline\hline
\end{tabular}
\label{tab:modes}
\end{table}

In the left panel of Fig. \ref{fig:delta}, we show the results for the ``fluid" model with the solid line and for the ``Sp" model with the dotted line. From this figure, one can observe the excitation of the shear ($s_i$-) and interface ($i_i$-) modes together with the fundamental ($f$-) and pressure ($p_i$-) modes in the ``Sp" model (see also the right panel), where $f$- and $p_1$-mode frequencies excited in the ``Sp" model are almost the same as those excited in the ``fluid" model. That is, the presence of the elasticity hardly affects the acoustic oscillations. In the middle panel, we show the results for the ``Sp" model with the dotted line and for the ``Sp+Cy" model with the solid line. From this result, one can observe that the $s_i$-mode frequencies in the ``Sp+Cy" model become smaller than those in the ``Sp" model. In the right panel, we show an enlarged view of the middle panel, where we also show the results for the ``realistic" model (SpCy+CHSH). From the right panel, one can observe that the $i_2$-mode frequency in the ``Sp+Cy" model is the same as that in the ``Sp" model, while the $i_1$-mode frequency strongly depends on the presence of the phase composed of cylindrical nuclei. In addition, for the ``realistic" model one can observe an additional mode, i.e., the $i_3$-mode, together with the $i_1$- and $i_2$-modes excited in the ``Sp+Cy" model. In particular, except for the excitation of the $i_3$-mode, we find that the eigenfrequencies excited in the ``realistic" model are the same as those in the ``Sp+Cy" model at least in the frequency domain shown in Fig. \ref{fig:delta}. This may be because the phase composed of cylindrical-hole and spherical-hole nuclei is quite narrow and the effect can not appear in the frequency domain considered here, as discussed below.

\begin{figure}[tbp]
\begin{center}
\includegraphics[scale=0.5]{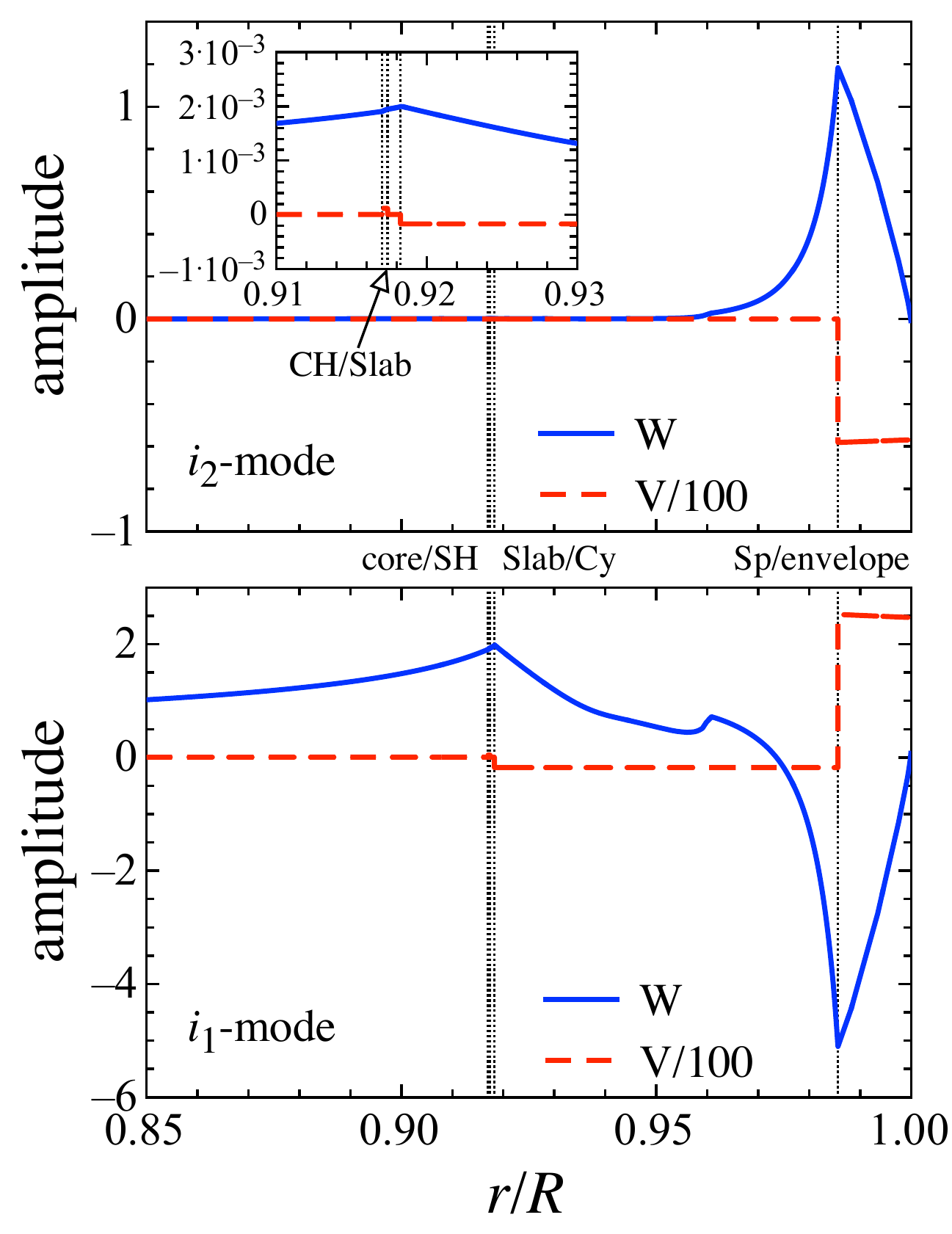}
\end{center}
\caption{
For a realistic stellar model with $1.4M_\odot$ and $12.4$ km using the EOS with $L=73.4$ MeV, the eigenfunctions of $W$ (solid lines) and $V$ (dashed lines) for the $i_2$-mode ($i_1$-mode) are shown in the top (bottom) panel. The vertical dotted lines from right to left denote the boundary between the crust and envelop (Sp/envelope), the boundary between the phases composed of cylindrical and slab-like nuclei (Slab/Cy), the boundary between the phases of slab-like and cylindrical-hole nuclei (CH/Slab), and the boundary between the phase composed of spherical-hole nuclei and core (core/SH). 
}
\label{fig:i-u36}
\end{figure}

\begin{figure}[tbp]
\begin{center}
\includegraphics[scale=0.5]{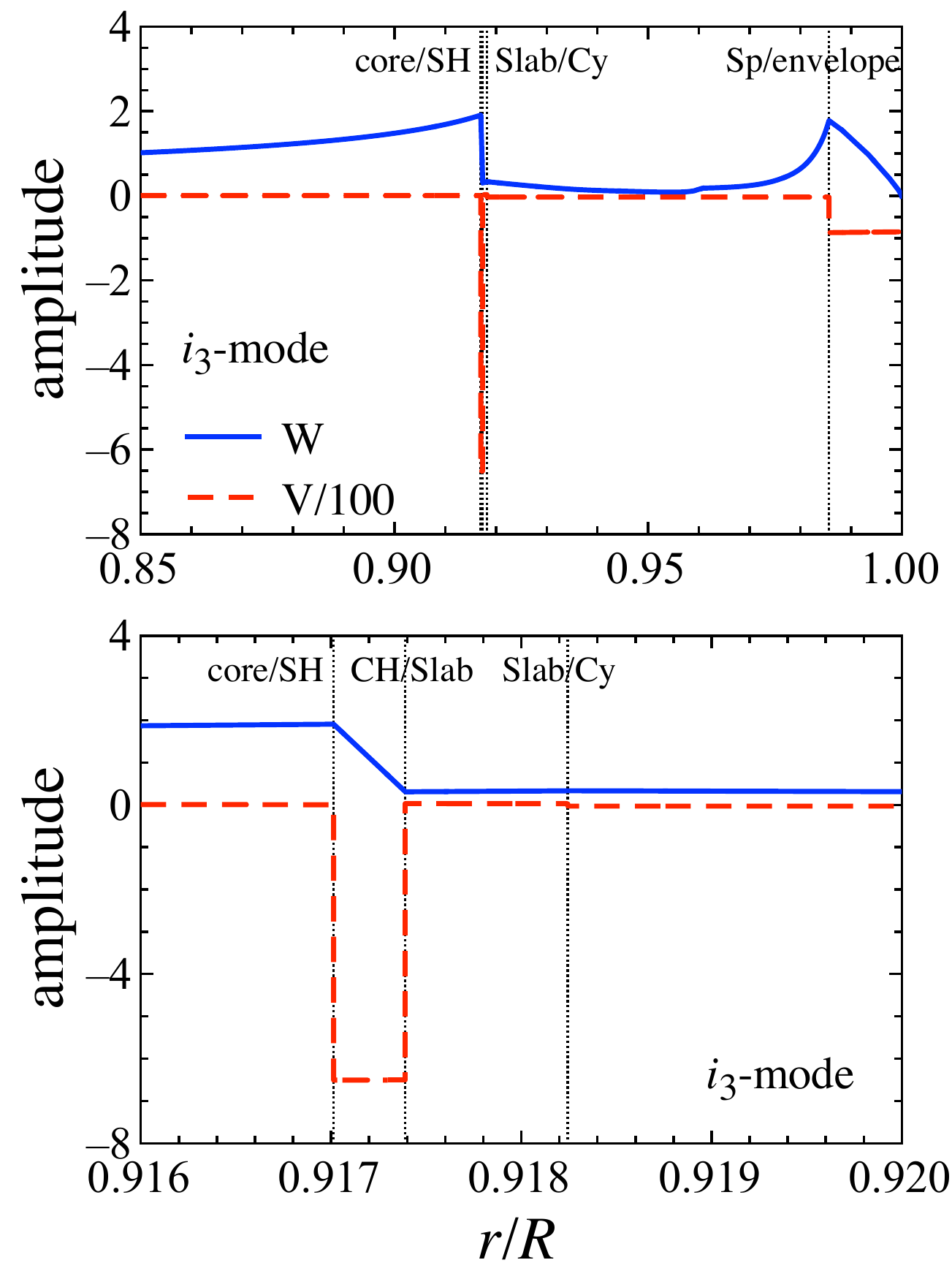}
\end{center}
\caption{
For a realistic stellar model with $1.4M_\odot$ and $12.4$ km using the EOS with $L=73.4$ MeV, the eigenfunctions of $W$ (solid lines) and $V$ (dashed lines) for the $i_3$-mode are shown in the top panel, while 
an enlarged view of the top panel is shown in the bottom panel. The meaning of the vertical lines is the same as in Fig. \ref{fig:i-u36}. }
\label{fig:i3-u36}
\end{figure}

The $i_i$-modes are the eigenmodes excited due to the presence of the interface between the phases with zero and non-zero elasticity \cite{PB05}. In this paper, we simply assign the $i_i$-modes in order from the highest to the lowest frequencies. That is, we have only two $i$-modes in the ``Sp+Cy" model as shown in the right panel of Fig. \ref{fig:delta}, because there are two interfaces, i.e., the interface between the envelope and crust and the interface between the phases composed of cylindrical and slab-like nuclei. In Fig. \ref{fig:i-u36} we show the amplitude of eigenfunctions, $W$ and $V$, for $i_2$-mode ($i_1$-mode) in the top (bottom) panel excited in the ``realistic" model, where the vertical dotted lines denote the boundary between the envelope and crust (Sp/envelope); the boundary between the phases of cylindrical and slab-like nuclei (Slab/Cy); the boundary between the phases of slab-like and cylindrical-hole nuclei (CH/Slab); and the boundary between the phase of spherical-hole nuclei and core (core/SH). From this figure, one can see that the $i_2$-mode is associated with the interface between the envelope and crust, while the $i_1$-mode is with the interface between the phases composed of cylindrical and slab-like nuclei. This may be a reason why the $i_2$-mode frequency in the ``Sp" model is the same as that in the ``Sp+Cy" model, as shown in the right panel of Fig. \ref{fig:delta}. Anyway, in the ``realistic" model, one can see the tiny effect of the non-zero elasticity in the phase composed of cylindrical-hole and spherical-hole nuclei in the amplitude of eigenfunctions, even for the $i_1$- and $i_2$-modes (e.g., see the inset in the top panel of Fig. \ref{fig:i-u36}). Moreover, in Fig. \ref{fig:i3-u36}, we show the amplitude of the $i_3$-mode in the top panel, while an enlarged view of the top panel is shown in the bottom panel. One can observe that the amplitude of the $i_3$-mode becomes dominant inside the region composed of the cylindrical-hole and spherical-hole nuclei. We also find that only three interface modes are excited in the ``realistic" model, even though four interfaces exist in the ``realistic" model, i.e., Sp/envelope, Slab/Cy, CH/Slab, and core/SH. This may come from the fact that the region composed of cylindrical-hole and spherical-hole nuclei is too narrow. In fact, if the elastic region becomes too narrow, the number of excited interface modes can become less than the number of interfaces, as shown in Appendix \ref{sec:appendix_4}. Furthermore, if one considers the neutron star model using the EOS with $L=42.6$ MeV, where the ratio of the thickness of the elastic region composed of cylindrical-hole and spherical-hole nuclei to the stellar radius is relatively larger than that considered in Fig. \ref{fig:delta} as shown in Table \ref{tab:EOS}, one can observe the $i_4$-mode together with the $i_3$-mode by introducing the elastic region composed of cylindrical-hole and spherical-hole nuclei (see Sec. \ref{sec:dep} for details).

\begin{figure}[tbp]
\begin{center}
\includegraphics[scale=0.5]{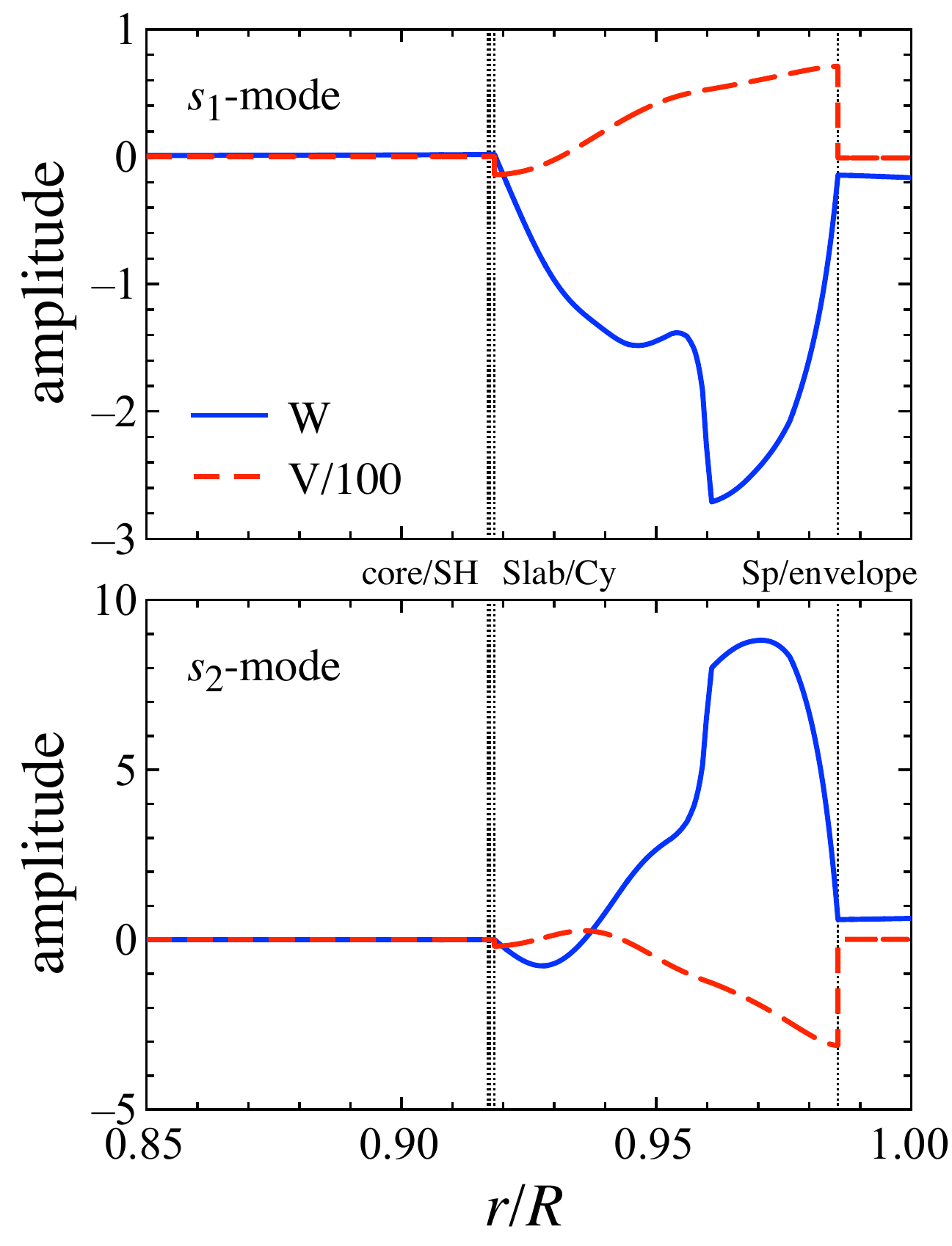}
\end{center}
\caption{
For a realistic stellar model with $1.4M_\odot$ and $12.4$ km using the EOS with $L=73.4$ MeV, the eigenfunctions of $W$ (solid lines) and $V$ (dashed lines) for the $s_1$-mode ($s_2$-mode) are shown in the top (bottom) panel. The meaning of the vertical lines is the same as in Fig. \ref{fig:i-u36}.
}
\label{fig:s-u36}
\end{figure}

\begin{figure}[tbp]
\begin{center}
\includegraphics[scale=0.5]{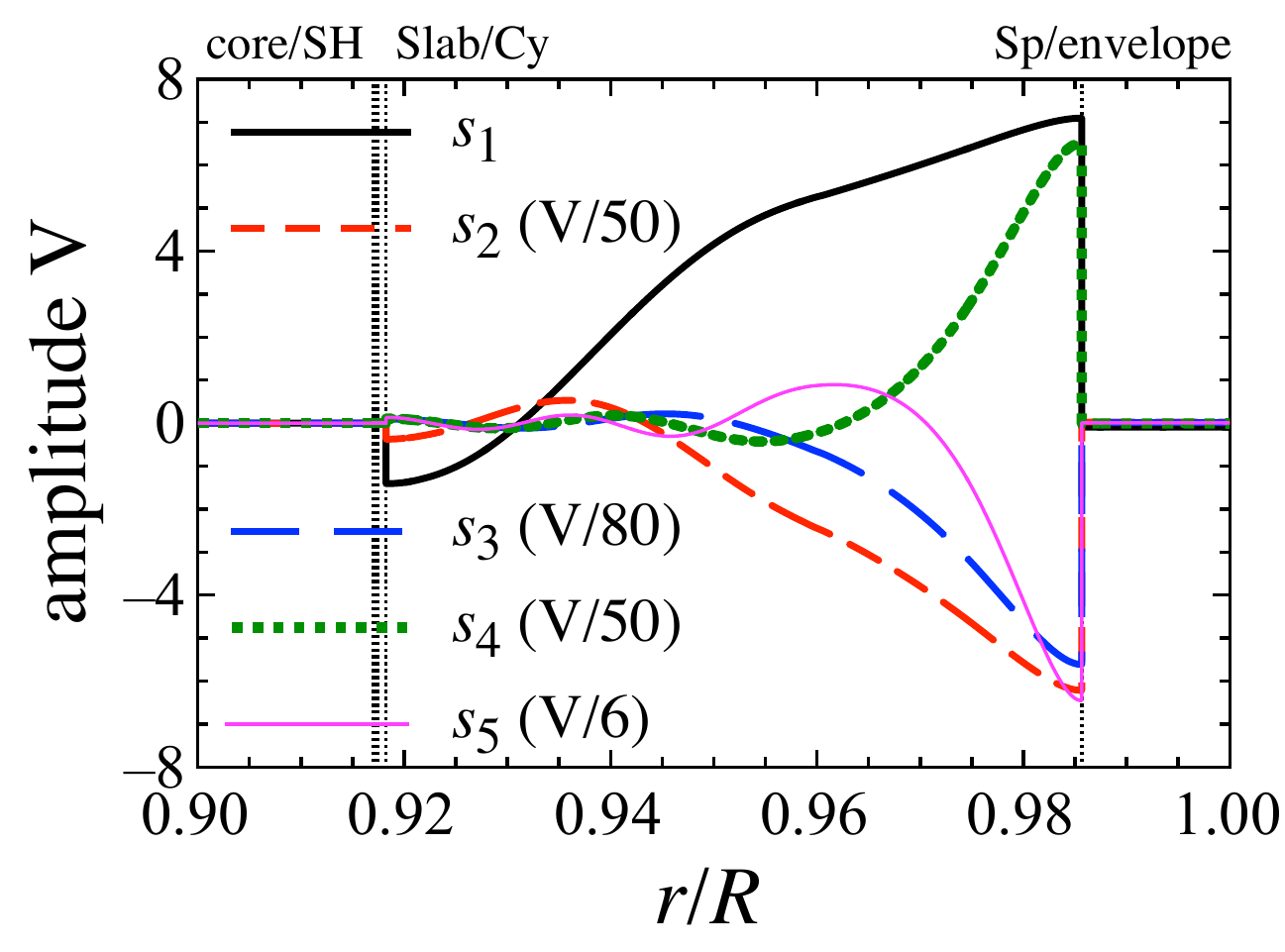}
\end{center}
\caption{
Eigenfunctions of $V$ for the $s_i$-modes with $i=1-5$. 
}
\label{fig:sv-u36}
\end{figure}

The $s_i$-modes are also the eigenmodes excited due to the presence of elasticity. Unlike the $i_i$-modes, the $s_i$-modes are basically confined inside the elastic region. In Fig. \ref{fig:s-u36}, we show the amplitude of the eigenfunction, $W$ and $V$, for the $s_1$-mode ($s_2$-mode) in the top (bottom) panel. From this figure, one can see that $W$ is continuous even at the boundaries between the anelastic and elastic regions owing to the junction condition (see in Appendix \ref{sec:appendix_3}), while $V$ is discontinuous at the boundaries. In addition, one can observe that the nodal number in the eigenfunction of the $s_i$-modes is equivalent to the subscript $i$. This behavior is easily observed in the amplitude of $V$, as shown in Fig. \ref{fig:sv-u36}, where we show the amplitude of V for the $s_i$-modes with $i=1-5$. So, the wavelength of the $s_i$-mode, $\lambda_i$, is roughly estimated as  
\begin{equation}
  \lambda_i \simeq 2\Delta R / i, \label{eq:lambda}
\end{equation}
where $\Delta R$ denotes the thickness of the elastic region, in which the shear modes are confined. Then, the corresponding frequencies are also estimated as
\begin{equation}
  f_i \approx v_s/ \lambda_i, \label{eq:fn}
\end{equation}
where $v_s\approx (\mu/\epsilon)^{1/2}$ denotes the shear velocity \cite{KHA2015}. With this simple estimation, one may understand why the $s_i$-mode frequencies excited in the ``realistic" model are the same as those in the ``Sp+Cy" model as discussed in Fig. \ref{fig:delta}. That is, the $s_i$-mode frequencies excited in the phase of cylindrical-hole and spherical-hole nuclei must be much higher (maybe more than 100 times higher) than the $s_i$-mode frequencies excited in the phase of spherical and cylindrical nuclei, because $\Delta R$ for the phase of cylindrical-hole and spherical-hole nuclei, $\Delta R_{\rm CHSH}$, is much thinner than $\Delta R$ for the phase of spherical and cylindrical nuclei, $\Delta R_{\rm SpCy}$, as shown in Table \ref{tab:EOS}, i.e., $\Delta R_{\rm SpCy}/\Delta R_{\rm CHSH}=179$.

\section{Dependence on the neutron star properties}
\label{sec:dep}

\begin{figure*}[tbp]
\begin{center}
\includegraphics[scale=0.5]{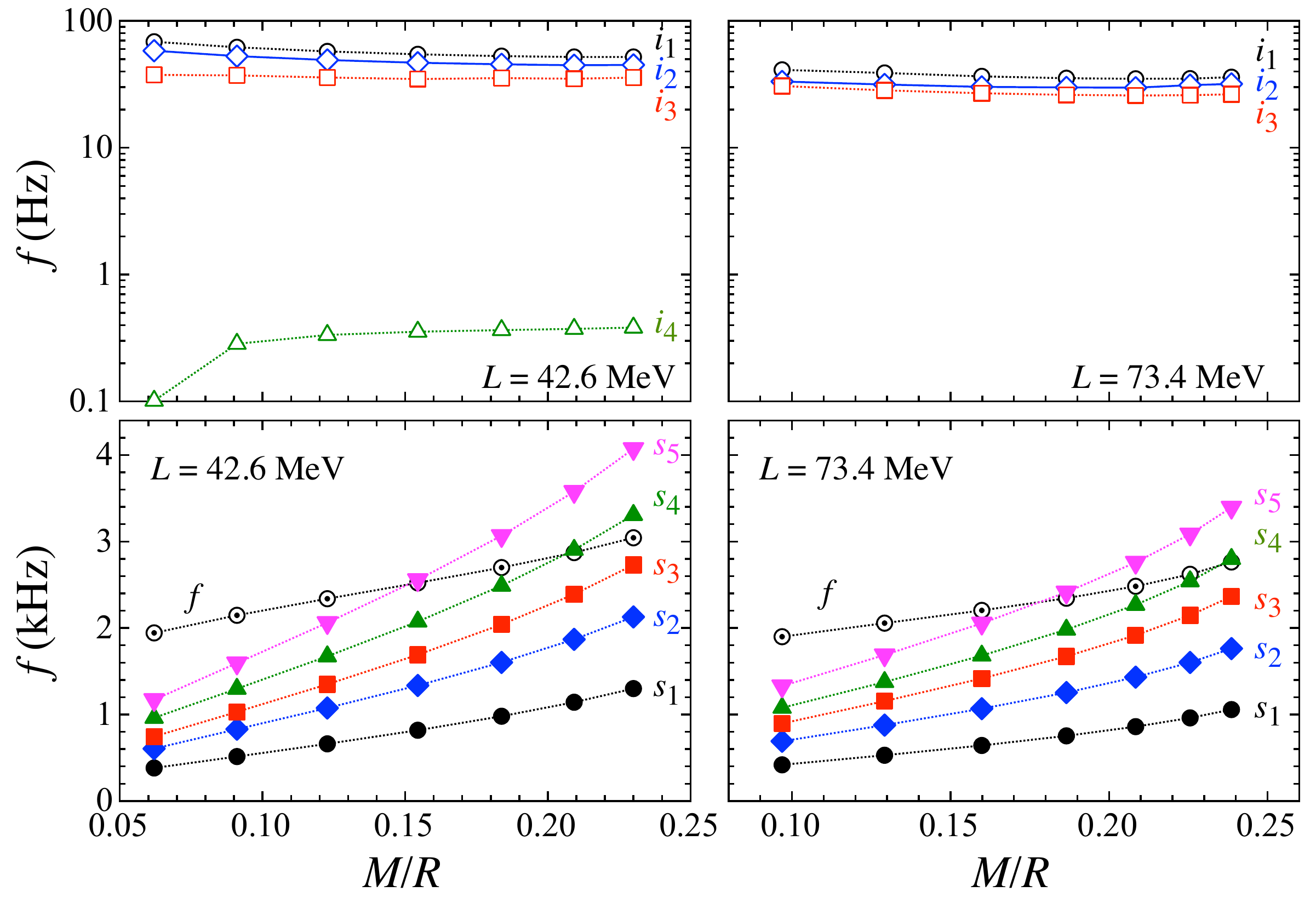}
\end{center}
\caption{
Eigenfrequencies of the $i_i$-modes (top panels) and the $s_i$- and $f$-modes (bottom panels) are shown as a function of the stellar compactness, $M/R$, for the neutron star models constructed using the EOSs with $L=42.6$ (left panels) and $73.4$ MeV (right panels).
}
\label{fig:y-MR}
\end{figure*}

First, in Fig. \ref{fig:y-MR}, we show the eigenfrequencies of the $i$-, $s$-, and $f$-modes as a function of the stellar compactness for the neutron star models constructed using the EOSs with $L=42.6$ and $73.4$ MeV. From this figure, one can observe that the $i$-mode frequencies weakly depend on the stellar compactness, while the $s$-mode frequencies monotonically increase with the stellar compactness. This is because the ratio of the thickness of the elastic region to the stellar radius decreases as the stellar compactness increases \cite{SIO2017}, which leads to the increase of the $s$-mode frequencies as discussed with Eqs. (\ref{eq:lambda}) and (\ref{eq:fn}). In addition, as mentioned before, since the thickness of the elastic region composed of the cylindrical-hole and spherical-hole nuclei for the neutron star model constructed with $L=42.6$ MeV is relatively larger than that with $L=73.4$ MeV, one can observe the $i_4$-mode together with $i_1$-, $i_2$-, and $i_3$-modes in the stellar model with $L=42.6$ MeV. 

Next, we examine how the frequencies of the $i$- and $s$-modes depend on the neutron star properties. In particular, since the nuclear properties in the core region (or in a higher-density region) are quite uncertain, we examine the frequencies of the $i$- and $s$-modes by changing the stiffness of the EOS in a higher-density region. For this purpose, in addition to the original OI-EOSs listed in Table \ref{tab:EOS}, we simply consider that the EOS for a lower density region of $\varepsilon\le\varepsilon_t$, i.e., OI-EOSs, is connected to the one-parameter EOS characterized by $\alpha$ for a higher density region of $\varepsilon\ge\varepsilon_t$, i.e.,
\begin{equation}
  p = \alpha(\varepsilon - \varepsilon_t) + p_t, \label{eq:EOS}
\end{equation}
where $p_t$ is given from the EOS for a lower-density region with $\varepsilon=\varepsilon_t$ and $\alpha$ is associated with the sound velocity, $c_s$, as $c_s^2=\alpha$ \cite{Sotani17}. In this study, we especially adopt that $\epsilon_t$ is equivalent to twice the saturation density, focusing on the value of $\alpha$ in the range of $1/3\le\alpha\le 1$.

\begin{figure}[tbp]
\begin{center}
\includegraphics[scale=0.5]{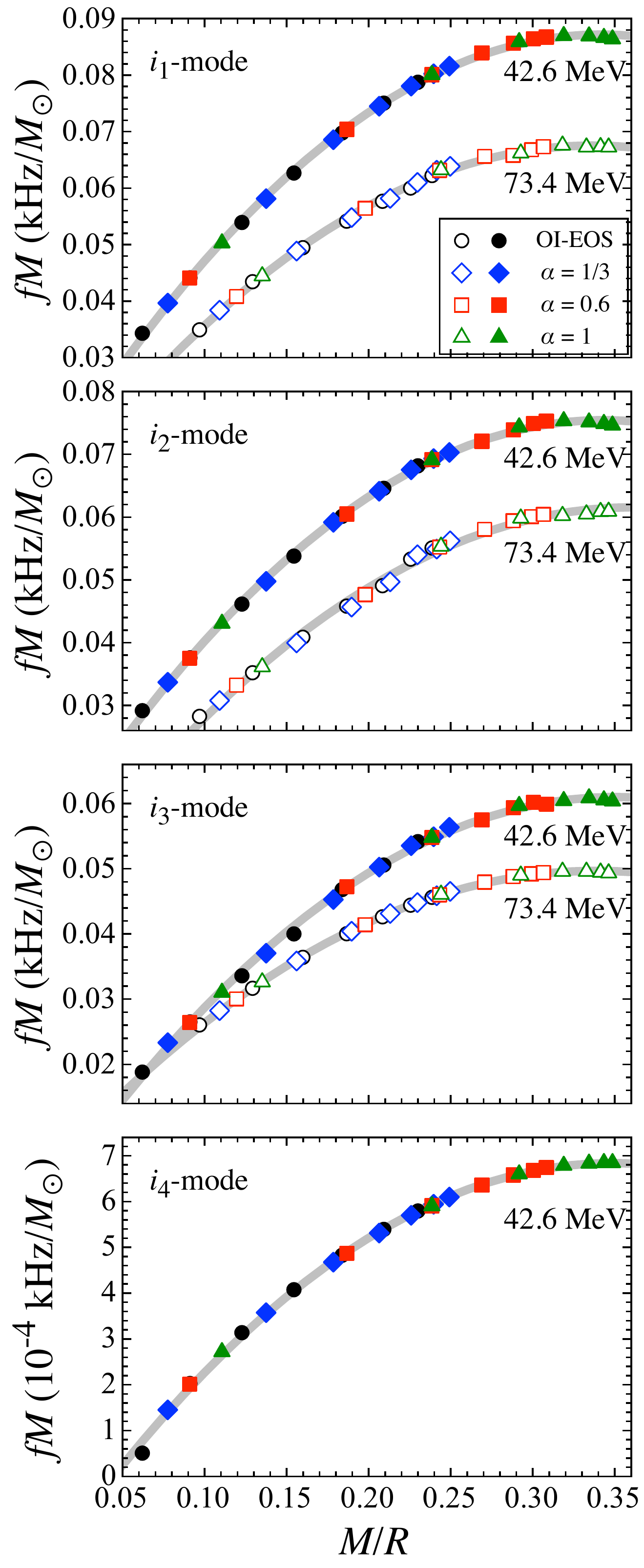}
\end{center}
\caption{
The $i_i$-mode frequencies multiplied by the stellar mass are shown as a function of the stellar compactness. The filled and open marks denote the results for the neutron star model using the EOSs with $L=42.6$ and $73.4$ MeV, respectively. The circles denote the results for the neutron star models with the original OI-EOSs, while the diamonds, squares, and triangles denote the results for the neutron star models with the OI-EOS connected to the EOS characterized by $\alpha=1/3$, 0.6, and 1, respectively. The solid lines denote the fitting lines using the formula given by Eq. (\ref{eq:fit_fi}).
}
\label{fig:fiM-MR}
\end{figure}

In practice, if one calculates the frequencies of $i$- and $s$-mode with this type of EOS, the frequencies depend on the value of $\alpha$. However, we find that the $i$-mode frequencies multiplied by the stellar mass, $fM$, can be expressed as a function of the stellar compactness almost independently of the value of $\alpha$ (or the stiffness in a higher density region inside the neutron stars), only depending on the stiffness of the curst EOS, as shown in Fig. \ref{fig:fiM-MR}. In this figure, the solid lines are the fitting lines given by the functional form as
\begin{equation}
  fM\ ({\rm kHz}/M_\odot) = a_0 + a_1(x/0.1) + a_2(x/0.1)^2, \label{eq:fit_fi}
\end{equation}
where $x$ denotes the stellar compactness, $M/R$, and $a_0$, $a_1$, and $a_2$ are adjusted coefficients.

\begin{figure}[tbp]
\begin{center}
\includegraphics[scale=0.5]{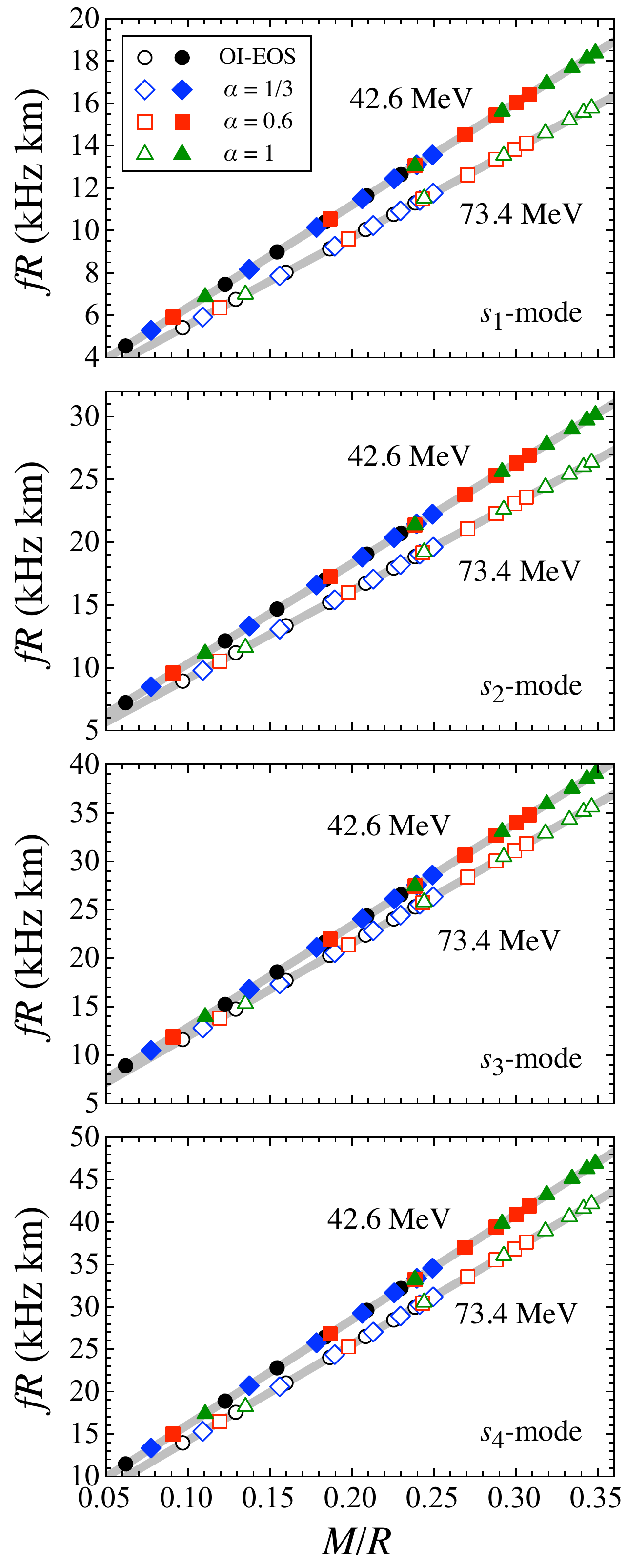}
\end{center}
\caption{
The $s$-mode frequencies multiplied by the stellar radius are shown as a function of the stellar compactness. The solid lines denote the fitting lines using the formula given by Eq. (\ref{eq:fit_fs}).
}
\label{fig:fsR-MR}
\end{figure}

In a similar way, we also find that the $s$-mode frequencies multiplied by the stellar radius, $fR$, can be expressed as a function of the stellar compactness almost independently of the value of $\alpha$, which depends only on the crust stiffness, as shown in Fig. \ref{fig:fsR-MR}. In this figure, the solid lines are fitting lines given by the functional form as
\begin{equation}
  fR\ ({\rm kHz\ km}) = b_0 + b_1(x/0.1), \label{eq:fit_fs}
\end{equation}
where $x$ denotes the stellar compactness, $M/R$, and $b_0$ and $b_1$ are adjusted coefficients.
Now, we find two different types of fitting formulae for the $i$- and $s$-mode frequencies. Thus, if one would simultaneously observe the $i$- and $s$-modes, one might extract the stellar mass and radius with the help of the constraint on the crust stiffness from the terrestrial experiments.

\section{Conclusion}
\label{sec:Conclusion}

We carefully examine the frequencies of the interface and shear oscillations, which are excited due to the presence of the elasticity, by considering the neutron star models with the pasta structures, i.e., cylindrical, slab-like, cylindrical-hole, and spherical-hole nuclei at the basis of the crust. We find that the shear mode frequencies excited in a realistic stellar model are basically the same as those in the neutron star model composed of only spherical and cylindrical nuclei, if we focus only on the frequency range up to a few kHz. This is because the shear modes are only excited inside the elastic region, which leads to the feature that the frequencies are inversely proportional to the thickness of the elastic region, and the thickness of the elastic region composed of cylindrical-hole and spherical-hole nuclei is extremely thin. On the other hand, the interface mode frequencies strongly depend on the elastic region composed of cylindrical-hole and spherical-hole nuclei. We find that the number of the interface mode frequencies depends on the thickness of the elastic region composed of cylindrical-hole and spherical-hole nuclei (or the value of the slop parameter $L$). In addition, we find the empirical relations for the interface mode frequencies multiplied by the stellar mass and for the shear mode frequencies multiplied by the stellar radius as a function of the stellar compactness, which is almost independent of the stiffness in a higher-density region inside the neutron stars, once one selects the crust equation of state. Via our empirical relations, if one would simultaneously observe the interface and shear mode oscillations from a neutron star, one might extract the stellar mass and radius with the help of the constraint on the crust stiffness obtained from the terrestrial experiments.

\acknowledgments
We are grateful to Shijun Yoshida, Christian J. Kr\"{u}ger, and Akira Harada for their valuable comments. This work is supported in part by Japan Society for the Promotion of Science (JSPS) KAKENHI Grant Numbers 
JP19KK0354  
and
JP21H01088,  
and by Pioneering Program of RIKEN for Evolution of Matter in the Universe (r-EMU).

\appendix
\section{Perturbation equations inside the elastic region}   
\label{sec:appendix_1}

The perturbation equations are derived from the linearized energy-momentum conservation laws:
\begin{align}
  r\frac{dz_1}{dr} =& -\left(1+\frac{2\alpha_2}{\alpha_3} + U_2\right)z_1 + \frac{1}{\alpha_3}z_2
     + \frac{\alpha_2}{\alpha_3}\ell(\ell+1)z_3, \label{eq:dz1} \\
  r\frac{dz_2}{dr} =& \left[\left(-3-U_2+U_1-e^{2\Lambda}c_1\bar\sigma^2\right)U_3 
     + \frac{4\alpha_1}{\alpha_3}\left(3\alpha_2 + 2\alpha_1\right)\right]z_1   \nonumber \\
     &+\left(U_4 - \frac{4\alpha_1}{\alpha_3}\right)z_2 
     + \left[U_3 - 2\alpha_1\left(1+\frac{2\alpha_2}{\alpha_3}\right)\right]\ell(\ell+1)z_3 \nonumber \\
     &+ e^{2\Lambda}\ell(\ell+1)z_4, \label{eq:dz2} \\
  r\frac{dz_3}{dr} =& -e^{2\Lambda}z_1 + \frac{1}{\alpha_1}e^{2\Lambda}z_4, \label{eq:dz3} \\
  r\frac{dz_4}{dr} =& \left(U_3 - 6\Gamma \frac{\alpha_1}{\alpha_3}\right)z_1 
     -\frac{\alpha_2}{\alpha_3}z_2 \nonumber \\
     &-\left[c_1\bar\sigma^2U_3 + 2\alpha_1 - \frac{2\alpha_1}{\alpha_3}\left(\alpha_2 
     + \alpha_3\right)\ell(\ell+1)\right]z_3  \nonumber \\
     &- \left(3+U_2-U_4\right)z_4, \label{eq:dz4}
\end{align}
where the variables, $z_i$ for $i=1-4$, are defined as
\begin{align}
  z_1 =& W, \\
  z_2 =& 2\alpha_1e^{-\Lambda}\frac{d}{dr}\left(re^{\Lambda}W\right) 
     + \left(\Gamma - \frac{2\alpha_1}{3}\right)
     \left[\frac{1}{r^2}e^{-\Lambda}\frac{d}{dr}\left(r^3e^{\Lambda}W\right)
     - \ell(\ell+1)V\right],  \\
  z_3 =& V,  \\
  z_4 =& \alpha_1\left(e^{-2\Lambda}r\frac{dV}{dr} + W\right).
\end{align}
We note that the variables $z_2$ and $z_4$ are proportional to the radial and transverse tractions \cite{MHH88}.
The various quantities in the coefficients are defined as
\begin{align}
 \alpha_1 =& \frac{\mu}{p}, \\
 \alpha_2 =& \Gamma - \frac{2\alpha_1}{3}, \\
 \alpha_3 =& \Gamma + \frac{4\alpha_1}{3}, \\ 
 U_1 =& \left(\frac{d\Phi}{dr}\right)^{-1}\frac{d}{dr}\left(r\frac{d\Phi}{dr}\right), \\
 U_2 =& r\frac{d\Lambda}{dr}, \\
 U_3 =& \left(1 + \frac{\epsilon}{p}\right)r\frac{d\Phi}{dr}, \\
 U_4 =& \frac{r\epsilon}{p}\frac{d\Phi}{dr}, \\
 c_1 =& \frac{M}{R^3}re^{-2\Phi}\left(\frac{d\Phi}{dr}\right)^{-1},
\end{align}
and $M$, $R$, and $\bar\sigma$ are the stellar mass, radius, and $\bar\sigma\equiv \sigma (R^3/M)^{1/2}$. 

\section{Perturbation equations in the fluid region}   
\label{sec:appendix_2}

One can derive the perturbation equations in the fluid region, where $\mu=0$:
\begin{align}
  r\frac{dy_1}{dr} =& -\left(3-\frac{U_3}{\Gamma} + U_2\right)y_1 
     - \left(\frac{U_3}{\Gamma}-\frac{\ell(\ell+1)}{c_1\bar\sigma^2}\right)y_2, \label{eq:dy1} \\
  r\frac{dy_2}{dr} =& \left(e^{2\Lambda}c_1\bar\sigma^2 + rA_r\right)y_1 
     - \left(U_1+rA_r\right)y_2, \label{eq:dy2}
\end{align}
where the variables, $y_1$ and $y_2$, are defined as
\begin{align}
  y_1 =& W, \\
  y_2 =& \left(r\frac{d\Phi}{dr}\right)^{-1}H = c_1\bar\sigma^2V.
\end{align}

\section{Boundary and junction conditions}   
\label{sec:appendix_3}

The boundary condition at the stellar surface is that the Lagrangian perturbation of pressure should be zero, i.e., $\Delta p=0$, which is expressed as
\begin{equation}
  y_1-y_2=0,
\end{equation}
while the boundary condition at the center is the regularity condition, such as
\begin{equation}
   c_1\bar\sigma^2y_1 - \ell y_2 =0.
\end{equation}
On the other hand, the junction conditions at the interface are the continuity of the radial displacement, the radial and transverse tractions, and the Lagrangian perturbation of pressure. In practice, the junction conditions at the interface between the elastic and fluid regions \cite{Finn90,SL02} are given as
\begin{align}
  z_1 =& y_1, \\
  z_2 =& U_3(y_1-y_2), \\
  z_4 =& 0.
\end{align}
In a similar way, the junction conditions at the interface, where the non-zero shear modulus becomes discontinuity, e.g., the interface between the phases composed of spherical and cylindrical nuclei, are the continuity of the variables $z_i$ for $i=1-4$. Finally, one can set that $y_1=1$ at the stellar surface (or at the stellar center), as a normalization condition in the linear perturbation system.


\section{Dependence on the thickness of elastic region}   
\label{sec:appendix_4}

\begin{figure}[tbp]
\begin{center}
\includegraphics[scale=0.5]{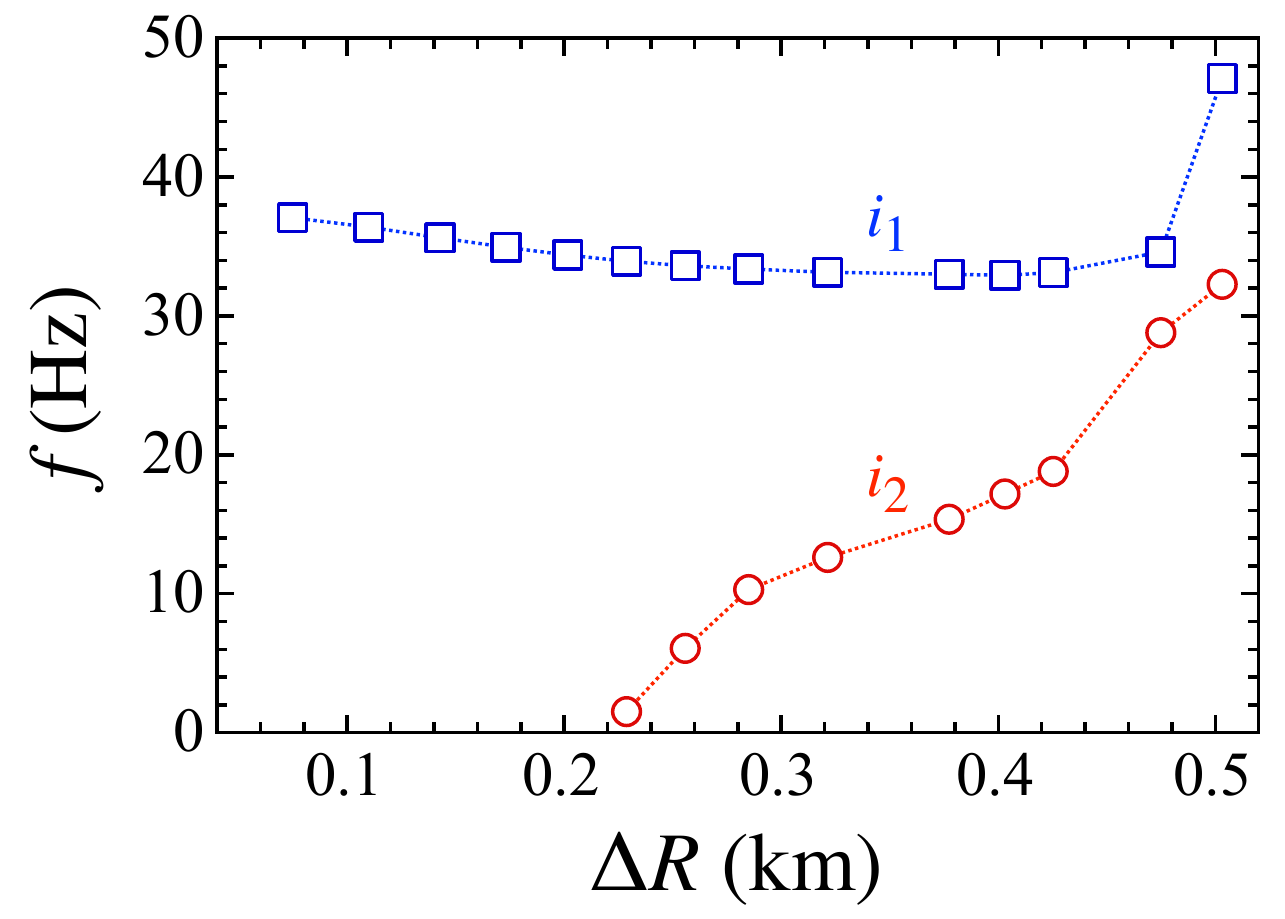}
\end{center}
\caption{
The frequencies of the $i_1$- and $i_2$-modes excited due to the presence of the elastic region composed of spherical and cylindrical nuclei are shown as a function of the thickness of the elastic region, $\Delta R$, by artificially increasing the density between the phase composed of spherical nuclei and envelope, for the neutron star model with $1.4M_\odot$ and 12.4 km, using the EOS with $L=73.4$ MeV.
}
\label{fig:fi-dR}
\end{figure}

In this appendix, we show how the frequencies of the interface modes depend on the thickness of an elastic region. To see this behavior, we especially consider the ``Sp+Cy" model as discussed in Sec. \ref{sec:modes} by artificially increasing the density between the phase composed of spherical nuclei and envelope from $10^{12}$ g/cm$^3$ up to $10^{14}$ g/cm$^3$. As this transition density ($\rho_{\rm Sp/En}$) increases, the thickness of the elastic region, $\Delta R$, decreases from $\Delta R=0.503$ to $0.075$ km. In Fig. \ref{fig:fi-dR}, the $i$-mode frequencies are shown as a function of $\Delta R$, where the $i_1$- and $i_2$-modes are the same meaning as discussed in Sec. \ref{sec:modes}. From this figure, one can observe that the frequency of the $i_1$-mode is almost independent of $\Delta R$ (or $\rho_{\rm Sp/En}$) with a lower value of $\Delta R$, while that of the $i_2$-mode strongly depends on $\Delta R$ (or $\rho_{\rm Sp/En}$). That is, as $\Delta R$ decreases (or as $\rho_{\rm Sp/En}$ increases), the frequency of the $i_2$-mode decreases and eventually disappears (or at least too small to be determined numerically). That is, even if two interfaces exist, only one interface mode can be excited in the stellar model with a much narrower elastic region.

\end{document}